\documentclass{webofc}
\usepackage{comment}
\usepackage[varg]{txfonts}   
\usepackage{hyperref}

\newcommand{\vb}[1]{\mathbf{#1}}
\newcommand{\pdv}[2][]{\frac{\partial{#1}}{\partial{#2}}}
\newcommand{\dd}[2][]{\mathrm d^{#1}{#2}\,}

\newcommand{\Conetwo}{\mathcal {C}^{1\leftrightarrow 2}}
\newcommand{\Ctwotwo}{\mathcal{ C}^{2\leftrightarrow 2}}

\newcommand{\nc}{N_{\mathrm{c}}}
\newcommand{\tauT}{\tau_{\mathrm{BMSS}}}
\newcommand{\tauR}{\tau_{\mathrm{R}}}
\newcommand{\qhat}{\hat q}

\begin{document}
\title{Limiting attractors in heavy-ion collisions---the interplay between bottom-up and hydrodynamic attractors}

\author{\firstname{Kirill} \lastname{Boguslavski}\inst{1}
    \and
        \firstname{Aleksi} \lastname{Kurkela}\inst{2}
             \and
        \firstname{Tuomas} \lastname{Lappi}\inst{3,4}
    \and
        \firstname{Florian} \lastname{Lindenbauer}\inst{1}\fnsep
        \thanks{Speaker, \email{florian.lindenbauer@tuwien.ac.at}}
    \and 
     \firstname{Jarkko} \lastname{Peuron}\inst{3,4}
}

\institute{Institute for Theoretical Physics, TU Wien, Wiedner Hauptstra{\ss}e 8-10, 1040 Vienna, Austria 
\and
           Faculty of Science and Technology, University of Stavanger, 4036 Stavanger, Norway 
\and
           Department of Physics, University of Jyväskylä,
P.O. Box 35, 40014 University of Jyväskylä, Finland
\and
Helsinki Institute of Physics, P.O. Box 64, 00014 University of Helsinki, Finland
          }

\abstract{
  In this contribution to the Quark Matter 2023 proceedings, we study the hydrodynamization process in heavy-ion collisions using QCD kinetic theory and introduce the new concept of limiting attractors.
  They are defined via an extrapolation of observables to vanishing and infinite couplings.
  We find that the pressure ratio exhibits both a hydrodynamic and a bottom-up limiting attractor, while the ratios of hard probes transport coefficients $\qhat^{zz}/\qhat^{yy}$ and $\kappa_T/\kappa_z$ are better described in terms of the new bottom-up limiting attractor.
}
\maketitle
\section{Introduction}
\label{intro}
We consider the initial stages in heavy-ion collisions that are governed by the evolution of the non-equilibrium quark-gluon plasma created therein.
Using  QCD kinetic theory, we focus on the system's approach to hydrodynamics, and establish the new concept of limiting attractors, which are defined as an extrapolation to vanishing and infinite coupling $\lambda$ at a fixed rescaled time. In particular, the bottom-up limiting attractor is associated with the time scale $\tauT$ and $\lambda\to 0$, whereas the hydrodynamic limiting attractor is obtained for the relaxation time scale $\tauR$ and extrapolation $\lambda\to\infty$, with the relevant times scales given by
\begin{align}
    \tauT = \alpha_s^{-13/5}/Q_s\;, && \tauR = \frac{4\pi\eta/s}{T}.\label{eq:timescales}
\end{align}
The bottom-up time scale $\tauT$ stems from the weak coupling picture of bottom-up thermalization \cite{Baier:2000sb}. There, the dynamics
consists of over- and under-occupied stages,
and $\tauT$ is the time at which thermalization occurs, and $Q_s$ is the saturation scale. 
The relaxation time $\tauR$ is motivated by the form of the pressure ratio in conformal first-order hydrodynamics
\begin{align}
    \frac{P_L}{P_T}=1-8\frac{\eta/s}{\tau T}=1-\frac{2}{\pi}\frac{\tau_R}{\tau},
\end{align}
which depends only on the ratio $\tau/\tauR$. Here, the only medium parameter is the shear viscosity $\eta$. The dimensionless ratio $\eta/s$ describes the interaction strength of the system, and $T$ denotes the temperature.
Thus, with time rescaled by $\tau_R$, 
a universal curve emerges at late times, the hydrodynamic attractor \cite{Heller:2015dha}.
Related attractors have also been found numerically in kinetic theory \cite{Almaalol:2020rnu} at fixed coupling, and for a variation of couplings \cite{Kurkela:2018wud}.

A natural question to ask is how to reconcile these two time scales from Eq.~\eqref{eq:timescales},
and which of these is more relevant for a given observable.
In particular, we are interested in transport coefficients of hard probes, which have recently received increased interest during the initial stages \cite{Ipp:2020nfu, Boguslavski:2020tqz, Carrington:2022bnv, Avramescu:2023qvv, Boguslavski:2023alu, Boguslavski:2023fdm, Du:2023izb}.
We address this question using QCD kinetic theory \cite{Arnold:2002zm} and perform simulations for a wide range of couplings $0.5\leq \lambda\leq 20$, and different initial conditions \eqref{eq:initial_cond} with varying initial anisotropy.

\section{Setup and kinetic theory}
\label{sec-kinetic-theory}
We describe the non-equilibrium plasma as a weakly interacting system of gluons.
In kinetic theory, the plasma is characterized by the distribution function $f(\vb p, \tau)$, which we evolve in time using the Boltzmann equation \cite{Arnold:2002zm}
\begin{align}
    \left(-\partial_\tau+\frac{p_z}{\tau}\pdv{p_z}\right)f(\vb p,\tau)=\Ctwotwo[f]+\Conetwo[f].
\end{align}
At leading order, the collision terms $\Ctwotwo$ and $\Conetwo$ are complicated functionals of the distribution itself. Our numerical approach and initial conditions follow  Ref.~\cite{Kurkela:2015qoa},
\begin{align}
    f(p_\perp,p_z,\tau=1/Q_s)=\frac{2A(\xi)\langle p_T\rangle}{\lambda\sqrt{p_\perp^2+(\xi p_z)^2}}\exp\left({-\frac{2(p_\perp^2+(\xi p_z)^2)}{3\langle p_T\rangle^2}}\right), &&\langle p_T\rangle =1.8\,Q_s, \label{eq:initial_cond}
\end{align}
where $\xi$ modifies the initial anisotropy and $A(\xi)$ is chosen such that the initial energy density is the same for different values of $\xi$.
Here, we assume homogeneity in the transverse plane, boost-invariance, and are interested in the mid-rapidity region.

The anisotropy of the plasma can be deduced from the difference of the longitudinal $P_L$ and transverse pressure $P_T$, which can be obtained from the energy-momentum tensor $T_{\mu\nu}$,
\begin{align}
    T^{\mu\nu}=2(\nc^2-1)\int\frac{\dd[3]{\vb p}}{(2\pi)^3}\frac{p^\mu p^\nu}{|\vb p|}f(\vb p), && P_L=T_{zz}, && P_T=T_{xx}=T_{yy},
\end{align}
where $p^0=|\vb p|$ and $\nc$ is the number of colors (3 for QCD). Another parameter is the occupancy of the hard sector $\langle pf\rangle/\langle p\rangle=\int\dd[3]{\vb p}|\vb p|(f(\vb p))^2/\int\dd[3]{\vb p}|\vb p|f(\vb p)$. Similar as in previous works \cite{Boguslavski:2023alu, Boguslavski:2023fdm}, we use the pressure ratio and occupancy to define time markers: The star marker is placed at occupancy $\langle pf \rangle/\langle p\rangle=\lambda$, the circle marker at minimum occupancy, and the triangle marker at the pressure ratio $P_T/P_L=2$, signaling an almost isotropic system.

\section{Results for pressure ratio and transport coefficients}

\begin{figure}
    \centering
    \centerline{
        \includegraphics[width=0.32\linewidth]{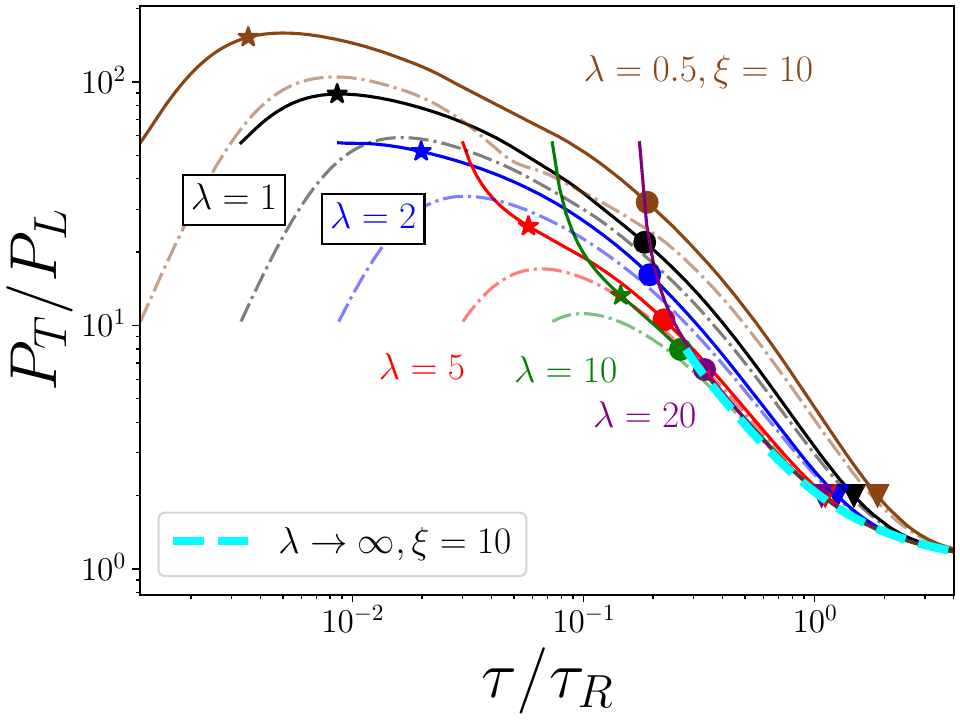}
        \includegraphics[width=0.32\linewidth]{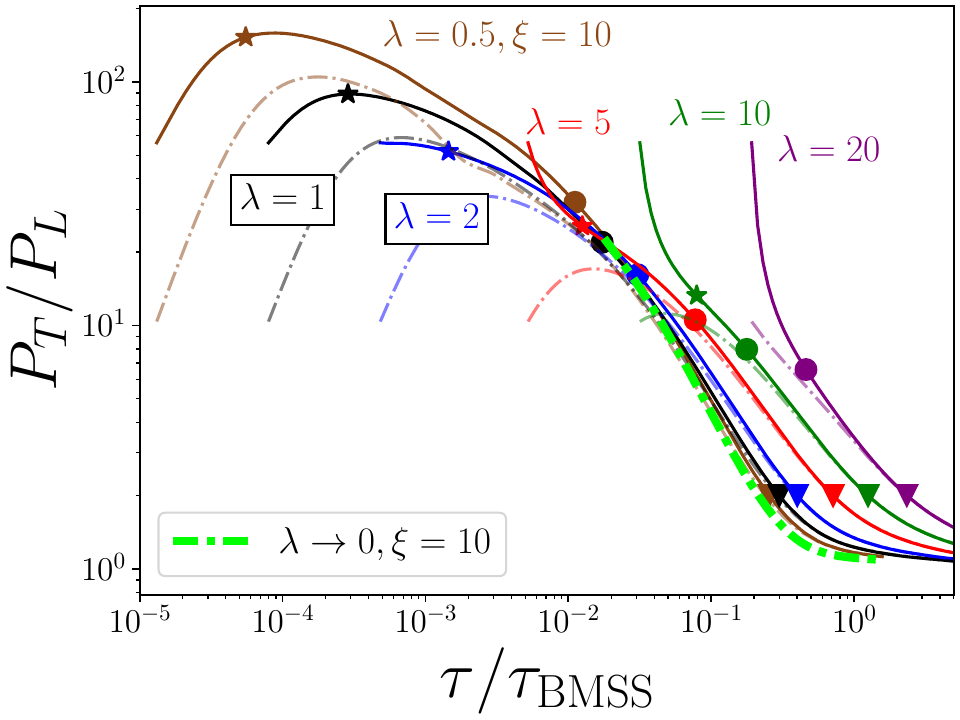}
        \includegraphics[width=0.32\linewidth]{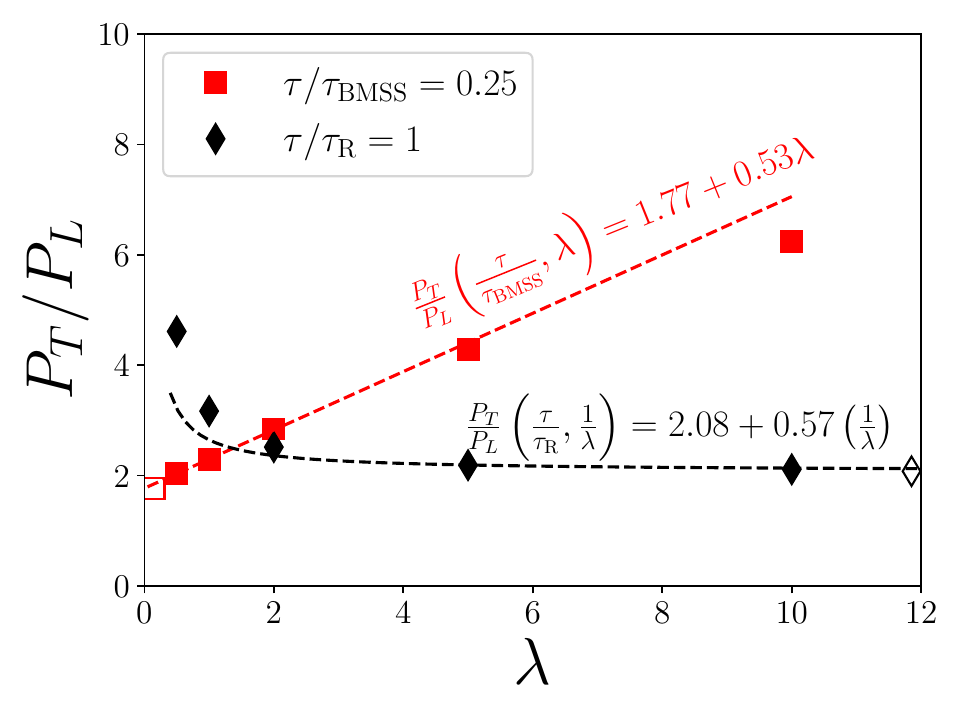}
    }
    \caption{\label{fig:pressure-ratio}Pressure ratio $P_T/P_L$ for simulations with different couplings (indicated by color, labeled in the plots) and initial conditions ($\xi=10$: solid lines, $\xi=4$: dash-dotted lines). In the \emph{left panel,} time is rescaled with $\tauR$, and we show the limiting hydrodynamic attractor $\lambda\to\infty$, while in the \emph{middle panel}, time is rescaled with $\tauT$ and we show the limiting bottom-up attractor $\lambda\to 0$. In the \emph{right panel,} we illustrate the extrapolation procedure to obtain the limiting attractors at the fixed times $\tau/\tauT=0.25$ and $\tau/\tauR=1$. The full markers show our data values, whereas the dashed lines illustrate our fitting functions. The empty square and diamond mark the values for $\lambda\to 0$ and $\lambda\to\infty$, respectively. Figures taken from \cite{limitingattractors}.}
\end{figure}

We present our results for the pressure ratio in Fig.~\ref{fig:pressure-ratio}, with the time variable rescaled according to both time scales of Eqs.~\eqref{eq:timescales}. The pressure ratios for the same coupling $\lambda$ but different initial conditions (different line styles) converge towards each other, signaling the onset of an attractor.
Additionally, when rescaled with the relaxation time $\tauR$ (left panel), the pressure ratio for different couplings $\lambda$ approach the limiting hydrodynamic attractor curve, which we obtain by an extrapolation of $1/\lambda\to 0$.  It is clearly visible that for small couplings, the approach to this universal curve happens at very late times.
In contrast, when the time is rescaled with $\tauT$ (middle panel), it is possible to extract a weak-coupling bottom-up limiting attractor by linear extrapolation $\lambda\to 0$. 
The extrapolation procedures are illustrated in the right panel, where the pressure ratio is shown for a fixed value of $\tau/\tauR=1$ as black diamonds, and for fixed $\tau/\tauT=0.25$ as red squares, together with the corresponding fits (dashed lines) and extrapolated values (transparent markers).

\begin{figure*}[t]
\sidecaption
\begin{minipage}[b]{0.7\textwidth}
\centerline{
    \includegraphics[width=0.49\linewidth]{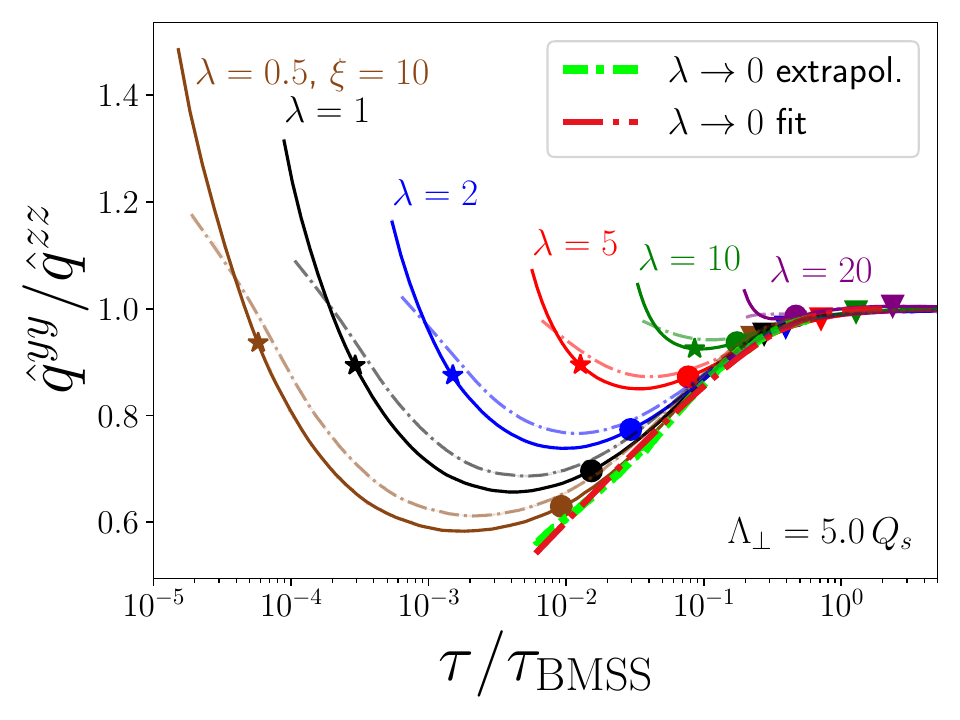}
    \includegraphics[width=0.49\linewidth]{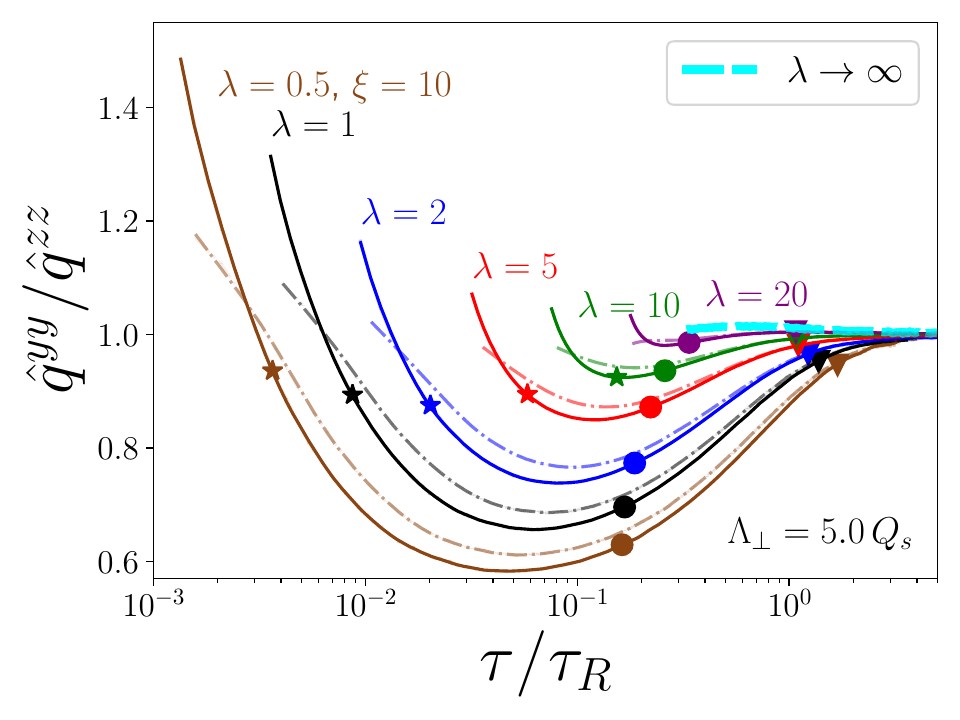}
}
\centerline{
    \includegraphics[width=0.49\linewidth]{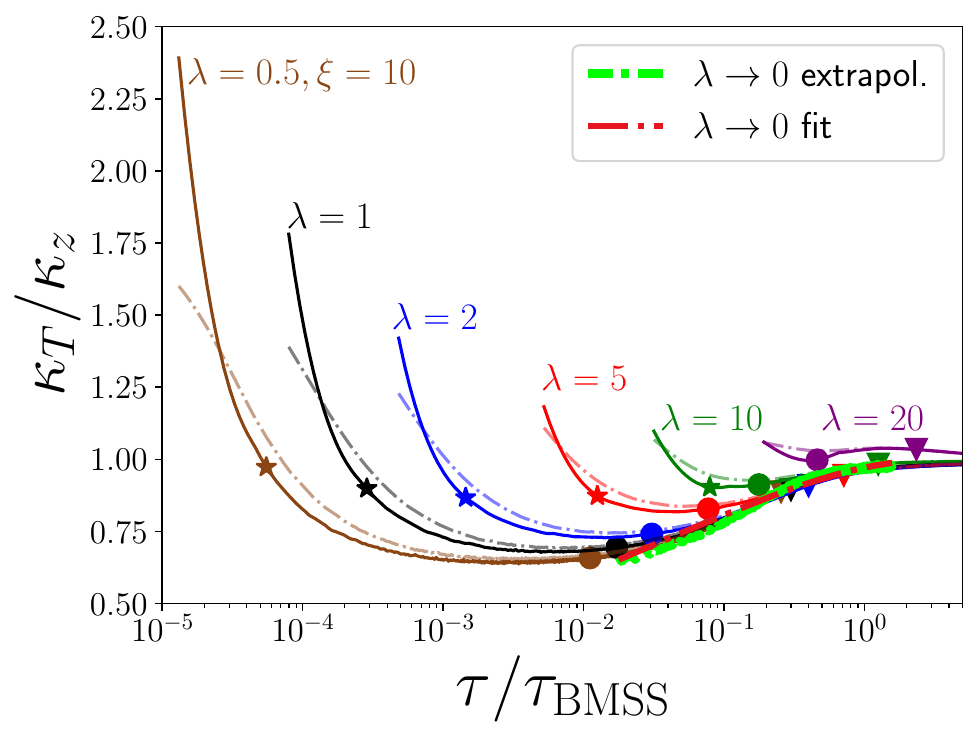}
    \includegraphics[width=0.49\linewidth]{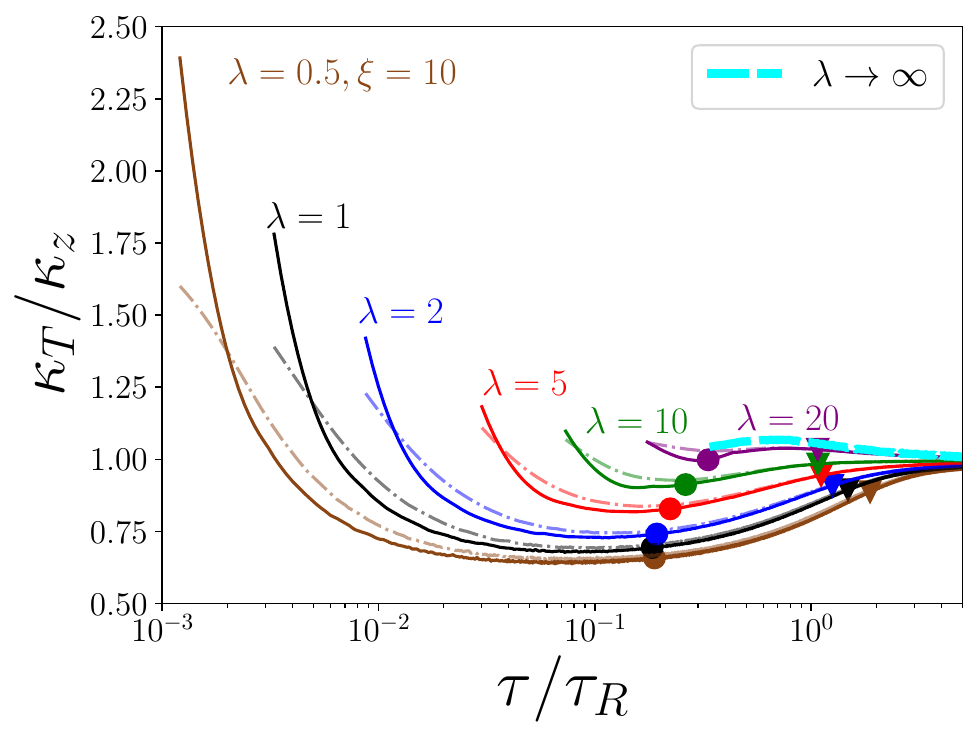}
}
\end{minipage}
\caption{\label{fig:transport_coeff}
\emph{Top panels}: Ratio of the jet quenching parameter $\qhat^{yy}/\qhat^{zz}$.
\emph{Bottom panels}: Ratio of the heavy quark diffusion coefficient $\kappa_T/\kappa_z$.
Each coupling is shown in a different color and initial conditions are distinguished by the line style.
\emph{Left column}: Time rescaled with $\tauT$, bottom-up limiting attractors are shown together with their parametrizations \eqref{eq:parametrization}.
\emph{Right column}: Time rescaled with $\tauR$, the hydrodynamic limiting attractors are included. Figures taken from \cite{limitingattractors}.
}
\end{figure*}

We now move on to study transport coefficients of hard probes that encode momentum broadening. In particular, we consider the momentum broadening of jets, described by the parameter $\hat q=\hat q^{yy}+\hat q^{zz}$, and of heavy quarks, encoded in $\kappa=(2\kappa_T+\kappa_z)/3$, given by \cite{Boguslavski:2023fdm, Boguslavski:2023waw}
\begin{align}
    \hat q^{ii}=\int\dd{\Gamma}\left(q^i\right)^2\left|\mathcal M\right|^2 f(\vb k)\left(1+f(\vb k')\right), && \kappa_i=\int\dd{\Gamma_k}\left(q^i\right)^2\left|\mathcal M_\kappa\right|^2f(\vb k)\left(1+f(\vb k')\right).
\end{align}
The anisotropy ratio of these transport coefficients is depicted in Fig.~\ref{fig:transport_coeff} with $\qhat^{yy}/\qhat^{zz}$ in the top, and $\kappa_T/\kappa_z$ in the bottom row. We first note the qualitatively similar evolution of these ratios, when rescaled with the same time: the bottom-up time scale $\tauT$ (left column), or the kinetic relaxation time $\tauR$ (right column).
While we can perform the extrapolation to infinite coupling at a fixed $\tau/\tauR$, the resulting hydrodynamic limiting attractor is only approached at very late times when it is already close to unity, as can be seen in the right column of Fig.~\ref{fig:transport_coeff}.
In contrast, one observes in the left column that the bottom-up limiting attractor
is approached at earlier times, even for larger values of the coupling. Thus, the bottom-up limiting attractor presents a more useful representation of these ratios for finite couplings. For convenience, we also provide a simple parametrization 
of this limiting curve for $\tau \gtrsim 0.01\,\tauT$, which we also include as red dash-dotted lines in the left column of Fig.~\ref{fig:transport_coeff}:
\begin{align}
    \label{eq:parametrization}
    R_{\qhat,\kappa}(\tau)=1+c_1^{\qhat,\kappa}\ln\left(1-e^{-c_2^{\qhat,\kappa}\tau/\tauT}\right) \qquad \text{with} \qquad \begin{cases}
        c_1^{\qhat}=0.12, \quad  c_2^{\qhat}=3.45\\
        c_1^{\kappa}=0.093, \quad c_2^{\kappa}=1.33
    \end{cases}
\end{align}

\section{Conclusions}
Using QCD kinetic theory simulations, we have established the new concept of limiting attractors in the initial stages in heavy-ion collisions. Both the hydrodynamic and bottom-up limiting attractors can be seen in the pressure ratio $P_T/P_L$, whereas for the ratios of hard probes transport coefficients $\qhat^{yy}/\qhat^{zz}$ and $\kappa_T/\kappa_z$ the bottom-up limiting attractor provides a significantly better description of the data.
This allows for a universal description of these ratios and thus a promising way to include pre-hydrodynamic effects for hard probes.

\section*{Acknowledgements}
This work is supported by the European Research Council, grant ERC-2015-CoG-681707, Academy of Finland (project 346324, and project 321840), and by the Austrian Science Fund (FWF) under project P34455 and W1252-N27. We wish to acknowledge the CSC - IT Center of Science, Finland, and the Vienna Scientific Cluster (project 71444) for computational resources.
\bibliography{bib}

\end{document}